# Spectral Evolution of Two High-Energy Gamma-Ray Bursts

Yuki Kaneko[1,2], Robert D. Preece[1,2], María Magdalena González[3,4], Brenda L. Dingus[4], and Michael S. Briggs[1,2]

[1]Department of Physics, University of Alabama in Huntsville, Huntsville, Alabama

[2]National Space Science and Technology Center, Huntsville, Alabama

[3]Department of Physics, University of Wisconsin at Madison, Madison, Wisconsin

[4]Los Alamos National Laboratory, Los Alamos, New Mexico

The prompt emission of the gamma-ray bursts is found to be very energetic, releasing $\sim 10^{51}$ ergs in a flash. However, their emission mechanism remains unclear and understanding their spectra is a key to determining the emission mechanism. Many GRB spectra have been analyzed in the sub-MeV energy band, and are usually well described with a smoothly broken power-law model. We present a spectral analysis of two bright bursts (GRB910503 and GRB930506), using BATSE and EGRET spectra that cover more than four decades of energy (30 keV – 200 MeV). Our results show time evolutions of spectral parameters (low-energy & high-energy photon indices and break energy) that are difficult to reconcile with a simple shock-acceleration model.

## INTRODUCTION

Gamma-ray bursts (GRBs) are among the most energetic phenomena in the universe and emit a tremendous amount of energy in seconds, primarily as gamma rays. Many GRBs of duration longer than a few seconds are followed by an afterglow of longer wavelengths, lasting days to months after the burst. Despite the numerous observations of GRBs and their afterglows, their creation mechanism and origins are still unclear. GRB spectra are non-thermal and continuous from a few keV to GeV; however, the distribution of the peak energy of the emitted power ($E_{peak}$) is found to be a narrow lognormal with a centroid value of ~250 keV [*Mallozzi et al.*, 1995; *Preece et al.*, 2000].

Many models have been suggested to explain the observed non-thermal spectra of GRBs. The most widely accepted picture is the synchrotron shock model (SSM). When shocks are formed, electrons are accelerated by the Fermi mechanism to a power-law energy distribution, $N(E_e) \propto E_e^{-p}$, and these highly relativistic electrons radiate synchrotron radiation due to the magnetic field behind the shock, producing the GRB. However, the SSM has some difficulties when confronted with the observational data [*Preece et al.*, 2000]. Studying the broadband energy spectra of GRBs is crucial to revealing the shock acceleration and gamma-ray emission mechanisms.

Here we present the results of broadband time-resolved spectral analysis of two spectrally hard GRBs.

## SPECTRAL ANALYSIS

*Instruments and Data Types*

The Burst and Transient Source Experiment (BATSE) on board the *Compton Gamma-Ray Observatory (CGRO)* observed 2704 GRBs in its 9-year lifetime (1991 − 2000). The BATSE observation provides the largest GRB database to date, with excellent time and energy resolution over the energy band 15 keV to ~10 MeV. Also on board the *CGRO* was the Energetic Gamma-Ray Experiment Telescope (EGRET), designed to observe gamma-ray sources in energies above 1 MeV. Some bright GRBs were observed with both BATSE and EGRET, providing spectra over a broader energy range.

BATSE was a collection of eight modules, each of which consisted of a Large Area Detector (LAD) and a Spectroscopy Detector (SD). Both are NaI(Tl) scintillation detectors coupled with photo multiplier tubes (PMTs). In this work, data from the brightest LAD (which varies from event to event, depending on the source direction) are used instead of the SD data due to the LAD's large effective area. There are several different data types for the LAD, of which two are used for this analysis − HERB (High Energy Resolution Burst) data and MER (Medium Energy Resolution) data. For both data types, the accumulation of the data began at the BATSE burst trigger. MER data are used when HERB data are not available or the HERB data are not complete, i.e., when the HERB data do not cover the entire duration of burst. Incomplete HERB data is common for bright events since HERB had a fixed memory space that could fill before the burst was over.

EGRET consisted of a spark chamber and a calorimeter (Total Absorption Shower Counter − TASC). The TASC was located at the bottom of EGRET and was made of a much larger NaI(Tl) scintillation crystal than were used in BATSE. Independently from the spark chamber events, the TASC observed a few dozen GRBs in its Burst mode, which was initiated by a BATSE trigger. In the Burst mode, spectra were accumulated in four commandable time intervals (normally 1, 2, 4, and 16 seconds). Each detector's characteristics are listed in Table 1.

*LAD-TASC Joint Spectral Analysis*

Combining the LAD and TASC data provides spectra that span 4 decades of energy (30 keV − 200 MeV). In general, GRB spectra are well fit with two power-laws joined smoothly at a break energy that is uniquely related to $E_{peak}$. As $E_{peak}$ approaches the upper limit of the BATSE passband, the BATSE data alone cannot adequately determine the high-energy power law index ($\beta$). Having TASC data along with BATSE data can extend the spectrum energy range up to 200 MeV, which may constrain $\beta$ as well as $E_{peak}$ values for the spectrally hard GRBs. The joint fit can also test the validity of the smoothly broken power-law model at higher energies that has been typically fitted using BATSE data.

The analysis was performed using the spectral analysis software RMFIT. To jointly fit time resolved spectra, LAD data were binned in time to match the TASC time bins. RMFIT employs forward-fitting procedures with one or more spectral models specified by users. In the actual fitting procedure, a multiplicative Effective Area Correction term was used because of uncertainties in the calculated effective areas of each detector. The goodness of fit is determined by $\chi^2$.

*GRB Spectral Model*

The photon model used in this analysis is an empirical "GRB" function, which consists of two power laws smoothly joined together [*Band et al.*, 1993]:

$$f(E) = A\left(\frac{E}{100\,\text{keV}}\right)^{\alpha} \exp\left(\frac{-(2+\alpha)E}{E_{peak}}\right)$$

if $E < (\alpha - \beta)\,[E_{peak}/(2 + \alpha)]$, and

$$f(E) = A\left(\frac{(\alpha - \beta)E_{peak}}{(2 + \alpha)100\,\text{keV}}\right)^{(\alpha - \beta)} \exp(\beta - \alpha)\left(\frac{E}{100\,\text{keV}}\right)^{\beta}$$

if $E \geq (\alpha - \beta)\,[E_{peak}/(2 + \alpha)]$;

where $A$ is the amplitude in photons $s^{-1}cm^{-2}keV^{-1}$, $E_{peak}$ is the peak energy of the power density spectrum, $\alpha$ is the low energy photon index, and $\beta$ is the high-energy photon index.

*The Events*

Two events, GRB910503 (BATSE trigger # 143) and GRB930506 (BATSE trigger # 2329), were selected for this analysis due to their brightness and their data availability. Lightcurves of these two bursts are shown in Figure 1. Both bursts are very hard and found to have fairly high $E_{peak}$ values, and therefore higher energy spectra are required to better constrain $E_{peak}$ and $\beta$ values.

## RESULTS AND DISCUSSION

Table 2 presents the best-fit spectral parameters for each event. The results clearly show the time evolution of the deduced photon spectra in each of the two events (see Figure 2).

*GRB910503 (Trigger # 143)*

It is evident that the spectra evolve from hard to soft with statistically significant changes in $\alpha$ and $\beta$ ($\Delta\alpha$ and $\Delta\beta$, respectively). Interestingly, we find $\Delta\alpha \sim \Delta\beta$. Moreover, the difference between $\alpha$ and $\beta$ (i.e., $\alpha - \beta \equiv \Delta s$) seems to remain approximately constant throughout the burst ($\Delta s \sim 1.6$). This value of $\Delta s$ is high compared with the average of $\sim 1.4$ or the most likely value of $\sim 1.0$ found by *Preece et al.* [2002] based on the analysis of 5500 time-resolved BATSE GRB spectra. For the first 2 time intervals, we find $\alpha > -2/3$ by about 4 $\sigma$ and 2 $\sigma$, respectively, which violates the synchrotron "line of death" predicted by the SSM [*Preece et al.* 1998].

*GRB930506 (Trigger # 2329)*

The spectra for this event do not seem to evolve from hard to soft, but rather $\beta$ stays constant while $\alpha$ and $E_{\text{peak}}$ evolve soft-hard-soft. In this case, since $\beta$ is clearly above $-2$, the fitted value for $E_{\text{peak}}$ is actually the break energy of the spectral model (where the high energy power law begins), and not the peak energy of the corresponding power density spectrum. This requires the existence of another spectral break (and thus the true $E_{\text{peak}}$) at an energy above the fitted $E_{\text{peak}}$ value.

The standard SSM involves optically-thin synchrotron radiation by energetic electrons that are left to radiate without further acceleration. The fact that the GRB spectra evolve on timescales much longer than the synchrotron cooling time may require an acceleration mechanism that is more complicated than those presumed in the SSM. Whatever allows the reacceleration of the electrons must somehow balance the very fast synchrotron cooling timescale. In addition, since $\beta$ is directly related to the power-law index of the shock-accelerated electron energy distribution, $p$, where $\beta = -(p+1)/2$, the changes in $\beta$ observed in GRB910503 may indicate change in $p$. Electrons accelerated by the Fermi mechanism are expected to have a power law distribution with $p \sim 2.2 - 2.3$ that is constant in time [*Gallant, Achterberg & Kirk*, 1999]. This also implies that $\beta < -2$ at all times, which is contradicted by the observations. The currently-standard SSM does not account for our results; therefore, the SSM needs modifications or a new shock-acceleration model of the GRB emission mechanism is required.

## REFERENCES

Band, D.L., et al. BATSE Observation of Gamma-Ray Burst Spectra I – Spectral Diversity, Astrophysical Journal, 413, pp. 281-292, 1993.

Gallant, Y.A., Achterberg, A. and Kirk, J.G., Particle Accelera-

tion at Ultra-Relativistic Shocks – Gamma-Ray Burst Afterglow Spectra and UHECRs, *Astronomy & Astrophysics Supplement Series,* 138, pp. 549-550, 1999.

Mallozzi, R.S., et al. The $\nu F_\nu$ Peak Energy Distributions of Gamma-Ray Bursts Observed by BATSE, Astrophysical Journal, 454, pp. 597-603, 1995.

Mészáros, P., Theories of Gamma-Ray Bursts, *Annual Review of Astronomy and Astrophysics*, 40, pp. 137-169, 2002.

Preece, R.D., et al. The Synchrotron Shock Model Confronts a "Line of Death" in the BATSE Gamma-Ray Burst Data, *Astrophysical Journal Letters,* 506, pp. 23-26, 1998.

Preece, R.D., et al. The BATSE Gamma-Ray Burst Spectral Catalog I – High Time Resolution Spectroscopy of Bright Bursts Using High Energy Resolution Data, *Astrophysical Journal Supplement Series,* 126, pp. 19-36, 2000.

Preece, R.D., et al. On the Consistency of Gamma-Ray Burst Spectral Indices with the Synchrotron Shock Model, *Astrophysical Journal,* 581, pp. 1248-1255, 2002.

___________

M.S. Briggs, Y. Kaneko and R.D. Preece, National Space Science and Technology Center, 320 Sparkman Drive, Huntsville, Alabama, 35805

B.L. Dingus and M.M. González, M.S. H803 P-23, Los Alamos National Laboratory, Los Alamos, New Mexico, 87545

**Table 1. Detector Characteristics**

| Detector | Energy Range (MeV) | Time Resolution (sec) | No. of Energy Channels |
|---|---|---|---|
| LAD | 0.03 – 2 | 0.128[a] (HERB) 0.016  (MER) | 128 (HERB) 16 (MER) |
| TASC | 1 – 200 | 1, 2, 4, 16[b] (BURST) | 256 |

[a] Minimum time resolution; increases by 64 ms increments

[b] Commandable

**Table 2. Best Fit Parameters**

| Time since trigger | 0 – 1 sec | 1 – 3 sec | 3 – 7 sec | 7 – 23 sec |
|---|---|---|---|---|
| GRB910503 (Trigger # 143 : EAC[c] = 0.66) | | | | |
| $A$ (ph s$^{-1}$cm$^{-2}$keV$^{-1}$) | 0.05 ± 0.001 | 0.32 ± 0.003 | 0.13 ± 0.001 | — |
| $E_{peak}$ (keV) | 1040 ± 74 | 727 ± 16 | 600 ± 15 | — |
| $\alpha$ | −0.51 ± 0.04 | −0.60 ± 0.01 | −0.91 ± 0.01 | — |
| $\beta$ | −2.03 ± 0.04 | −2.22 ± 0.02 | −2.60 ± 0.05 | — |
| $\alpha - \beta = \Delta$s | 1.52 | 1.62 | 1.69 | — |
| $\chi^2$/dof | 374/325 | 380/325 | 394/325 | — |
| GRB930506 (Trigger # 2329 : EAC[c] = 0.54) | | | | |
| $A$ (ph s$^{-1}$cm$^{-2}$keV$^{-1}$) | — | 0.04 ± 0.001 | 0.09 ± 0.0005 | 0.07 ± 0.0004 |
| $E_{peak}$ (keV) | — | 540 ± 54 | 1104 ± 41 | 871 ± 32 |
| $\alpha$ | — | −1.06 ± 0.04 | −0.90 ± 0.01 | −1.24 ± 0.01 |
| $\beta$ | — | −1.93 ± 0.06 | −1.91 ± 0.02 | −1.92 ± 0.02 |
| $\alpha - \beta = \Delta$s | — | 0.87 | 1.01 | 0.68 |
| $\chi^2$/dof | — | 205/212 | 293/212 | 272/212 |

[c] Effective Area Correction – Multiplicative term to normalize TASC to LAD

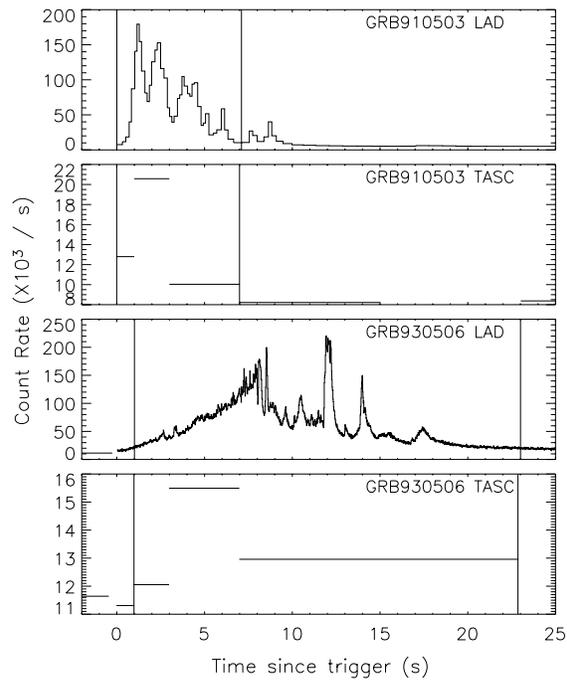

**Figure 1.** Time profiles of GRB 910503 (top two) and GRB 930506 (bottom two) as observed by BATSE and TASC. Time intervals used in the analysis are indicated.

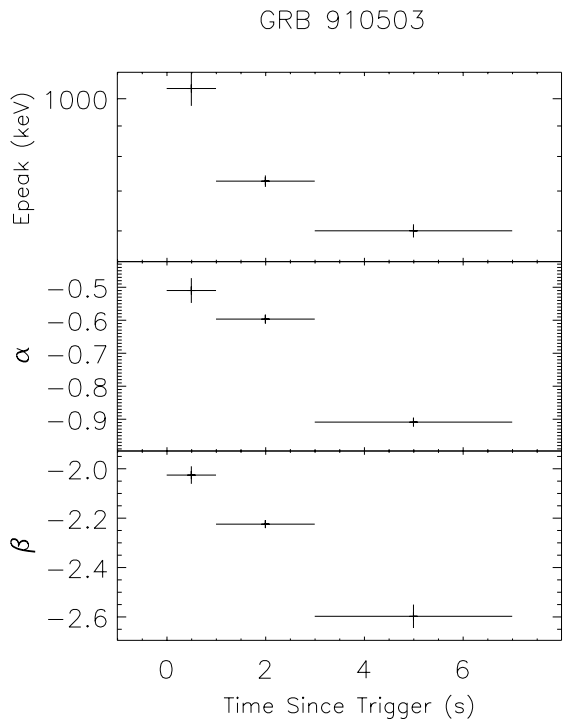

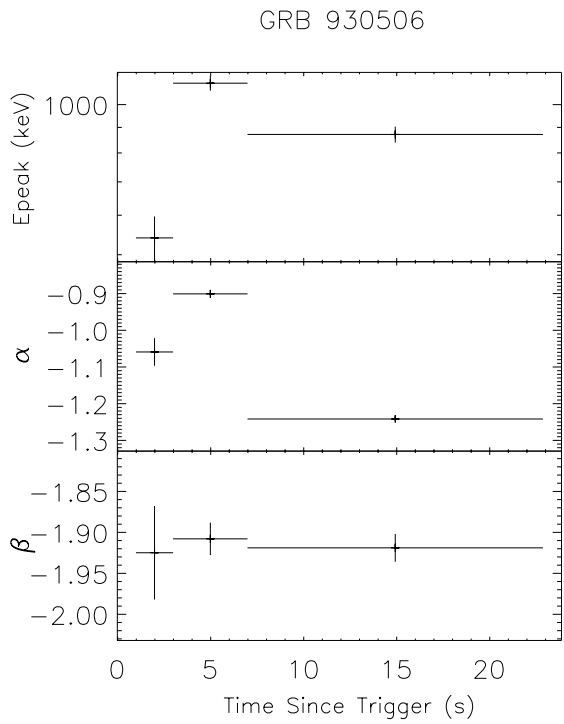

**Figure 2.** Spectral parameter evolutions in each burst. The values correspond to those in Table 2.